\documentclass[10pt,twocolumn,aps,pra,nobibnotes,showpacs,superscriptaddress,floatfix]{revtex4-1}

\usepackage{hyperref}
\usepackage[utf8]{inputenc}
\usepackage[T1]{fontenc}
\usepackage[sc,osf]{mathpazo}
\usepackage{amsmath, amsthm, amssymb,amsfonts}
\usepackage{graphicx}
\usepackage{dcolumn}
\usepackage{bm}
\usepackage{bbm}
\usepackage{mathtools}
\usepackage{comment}
\usepackage{color}

\def\RR{\mathbbm{R}}

\def\1{\mathbf{1}}
\def\0{\mathbf{0}}

\def\minimize{\textrm{minimize}}
\def\st{\textrm{subject to }}

\def\p{\mathbf{p}}

\def\h{\mathbf{h}}

\def\p{\mathbf{p}}
\def\h{\mathbf{h}}

\newcommand{\ket}[1]{| #1 \rangle}

\newcommand{\mean}[1]{\left\langle #1 \right\rangle}

\renewcommand{\rho}{\varrho}

\newcommand{\processnext}[1]{%
  \ifx\listfinish#1\empty\else\listact{#1}\expandafter\processnext\fi}

\newcommand{\figref}[1]{Fig.~\ref{#1}}

\begin{document}
\title{Entropic nonsignaling correlations}
\date{\today}

\author{Rafael Chaves}
\affiliation{Institute for Physics \& FDM, University of Freiburg, 79104 Freiburg, Germany}
\affiliation{Institute for Theoretical Physics, University of Cologne, 50937 Cologne, Germany}
\affiliation{International Institute of Physics, Universidade Federal do Rio Grande do Norte, 59070-405 Natal-RN, Brazil}
\author{Costantino Budroni}
\affiliation{Naturwissenschaftlich-Technische Fakult\"at, Universit\"at Siegen, Walter-Flex-Stra\ss{}e 3, 57068 Siegen, Germany}

\begin{abstract}
We introduce the concept of entropic nonsignaling  correlations, i.e., entropies arising from
probabilistic theories that are compatible with the fact that we cannot transmit information instantaneously. We characterize and show the relevance of these entropic correlations in a variety of different scenarios, ranging from typical Bell experiments to more refined descriptions such as bilocality and information causality. In particular, we apply the framework to derive the first entropic inequality testing genuine tripartite nonlocality in quantum systems of arbitrary dimension and also prove the first known monogamy relation for entropic Bell inequalities. Further, within the context of complex Bell networks, we show that entropic nonlocal correlations can be activated.
\end{abstract}

\maketitle

Quantum nonlocality---the fact that correlations obtained in quantum experiments performed
by distant parties are incompatible with local hidden variable (LHV) models
\cite{Bell1964}---brings to light an intriguing aspect of quantum mechanics (QM) and
relativistic causality \cite{popescu2014nonlocality}. QM is in accordance with the
nonsignaling (NS) principle, that is, local manipulations by an experimenter cannot
influence the measurement outcomes of other distant experimenters. However, as demonstrated
by Popescu and Rohrlich \cite{Popescu1994}, special relativity alone cannot single out
quantum mechanical correlations as there are theories, beyond QM,  also in agreement with
NS. This result not only has triggered the search for physically well-motivated principles
for quantum mechanics
\cite{Navascues2010glance,Navascues2014almost,Pawlowski2009,Fritz2013local,Cabello2013,Sainz2014}
but also has led to new insights about its limitations for information processing
\cite{Van1999nonlocality,Brassard2006limit,Chiribella2010,Almeida2010,Chaves2015a}.

Given the intrinsic statistical nature of QM, probabilities give a natural framework for
nonlocality. Indeed, Bell inequalities and NS relations are nothing other than constraints on
probabilities arising in a given theory, local and NS, respectively \cite{Brunner2014}.
Nevertheless, different approaches are possible
\cite{abramsky2011sheaf,Braunstein1988,Chaves2012}. In particular, in the
information-theoretic approach to nonlocality
\cite{Braunstein1988,Cerf1997,Chaves2012,FritzChaves2013,Chaves2014,
Dagomir2014,raeisi2015entropic} the basic objects are the Shannon entropies \cite{Yeung2008}
of the observed data.

The information-theoretic approach provides a novel and useful alternative for both
conceptual and technical reasons. First, entropy is a key concept in both classical and
quantum information theory, thus developing a framework that focuses on entropies rather than
probabilities leads to new insights and applications
\cite{Barnum2010entropy,Dahlsten2012tsirelson,Janzing2013,Chaves2014b,Henson2014,poh2015probing,Chaves2015entropy,janzing2015algorithmic}.
For instance, the celebrated principle of information causality \cite{Pawlowski2009} is
nothing other than an entropic inequality bounding the correlations that can be achieved by
imposing a certain causal structure to quantum mechanics \cite{Chaves2015a}. Second,
entropies allow for a much simpler and compact characterization of classical and quantum
correlations in a variety of scenarios \cite{Fritz2012,Chaves2014,Henson2014,Chaves2015a}.
In spite of that, as opposed to the usual probabilistic description, little is known about
entropic Bell inequalities beyond very simple cases and remarkably nothing is known about
the structure imposed by the nonsignaling principle on the entropies of measurement
outcomes.

In this letter, we aim to further develop the information-theoretic approach to nonlocality
and, in particular, to define the concept of entropic nonsignaling correlations, i.e., the
entropies compatible with the nonsignaling principle. We characterize NS entropic
correlations in a variety scenarios: from usual bipartite and tripartite, to genuine
multipartite nonlocality \cite{Svetlichny1987,Gallego2012,Pironio2013,Pironio2013},
bilocality \cite{Branciard2010}, and information causality \cite{Pawlowski2009}.
Our framework can also be employed to derive monogamy relations
\cite{Masanes2006,Pawlowski2009b} between entropic Bell inequalities.
Furthermore, our methods highlight the use of entropic NS correlations as a novel tool to derive Bell inequalities in scenarios otherwise intractable.

\textit{Marginal scenarios, local and NS correlations.---}
In a quantum experiment, only some of the relevant observables are jointly measurable;
hence, we face fundamental restrictions on the empirically accessible joint probability
distributions. This fact is encoded in the notion of a marginal scenario. Given $n$ random
variables $\left\{ X_1,\dots,X_n \right\}$, a marginal scenario $\mathcal{M}$ is defined as
$\mathcal{M}=\left\{ S_1,\dots, S_{\vert \mathcal{M} \vert} \right\}$, $S_i \subseteq \left\{ X_1,\dots,X_n \right\}$, such that for each $S_i$ a joint probability distribution
$P[(X_s)_{s\in S_i}]$ is accessible \cite{Chaves2012,FritzChaves2013}. Clearly, it is
sufficient to consider maximal subsets.

A typical example is a Bell experiment: two separated parties, Alice and Bob, at each run of 
the experiment can perform one of $m$ different measurements, labeled as $A_x$ and $B_y$, 
respectively, on their shares of a joint system. Their marginal scenario is, then, 
$\mathcal{M}_{\mathrm{Bell}}=\left\{\left\{A_x,B_y \right\}\right\}_{x,y=1,\ldots,m}$, 
corresponding to the probability distributions $\p_{\mathrm{obs}}=p(a_x,b_y)$ \cite{fnote1}---
where $a_x$ labels the outcome when measurement $x$ has been performed 
(similarly for $b$)---estimated from the statistical data. As shown by Fine \cite{Fine1982}, a 
LHV model for the data can be equivalently defined as a joint probability distribution 
$\p=p(a_1,\dots,a_{m},b_1,\dots,b_m)$. Hence, a set of marginals is called local if it is 
consistent with a single joint probability distribution for all measurements.
This, in turn, implies the existence of a joint entropy of all possible measurements 
$H_{A_1 \dots A_{m} B_1 \dots B_m}$ and all its marginals \cite{Braunstein1988}, where 
$H_{X}:= H(X):=-\sum_{x}p(x)\log_2 p(x)$ stands for the Shannon entropy. They can be 
represented as a {$2^{2m}$-dimensional} vector 
$\h=\left(H_{\emptyset}, H_{A_1},\dots, H_{A_1 \dots A_{m} B_1\dots B_m}) \right)$. Notice 
that $H_{\emptyset}$ is defined to be $0$, but it is convenient to include it to have a more 
compact representation of the constraints satisfied by the entropy vector (cf. Appendix \ref{app:cone}).

The difference between the probabilistic and entropic description solely relies on how we quantify correlations. A marginal probability distribution $\p_{\mathrm{obs}}$ is local if we can construct a well-defined joint probability distribution $\p$. Similarly, marginal entropies $\h_{\mathrm{obs}}=H(A_x,B_z)$ are local if a joint entropy and all its marginals $\h$, arising from a (nonunique) joint probability distribution, can be defined.
The existence of a well-defined joint description $\p$ imposes strict constraints---the famous Bell inequalities---on the empirically observable marginal correlations $\p_{\mathrm{obs}}$ \cite{Fine1982,Pitowsky1989,Pitowsky1991}. Similarly, marginal entropic correlations $\h_{\mathrm{obs}}$ admitting an extension to $\h$ also obey strict constraints. The closure set of well-defined entropy vectors $\h$ defines a convex cone $\Gamma_{\rm E}$, that is, if $\h$ and $\h^{\prime}$ are in $\Gamma_{\rm E}$ so are $p\h+(1-p)\h^{\prime}$, with $0 \leq p \leq 1$, and $\lambda \h$, with $\lambda\geq 0$ \cite{fnote3}.
An explicit characterization of $\Gamma_{\rm E}$ is yet to be found, however, an outer approximation, characterized by finitely many linear inequalities \cite{Yeung2008} or, equivalently, in terms of finitely many extremal rays (vectors defined up to a positive factor \cite{aliprantis2007cones}), is known: the Shannon cone $\Gamma_{\mathrm{Sh}}$. Such inequalities basically amount to the positivity of the conditional entropy, i.e. $H(A|B) := H(A,B)-H(B) \geq 0$, and the positivity of the conditional mutual information, i.e., $I(A:B|C) := H(A,C)+H(B,C)-H(A,B,C)-H(C) \geq 0$, for disjoint subsets of variables $A,B,C$ (see Appendix \ref{app:cone} for further details). Thus, in full analogy with the probabilistic case \cite{Budroni12}, entropic Bell inequalities can be understood as the constraints arising from the projection of $\Gamma_{\rm E}$ onto observable coordinates, that is, the projection of $\h$ into $\h_{\mathrm{obs}}$ defining the Bell entropic cone $\Gamma_{\mathrm{Bell}}$.

\begin{figure}[ht]
\vspace{0.6cm}
\center
\includegraphics[width=0.6\columnwidth]{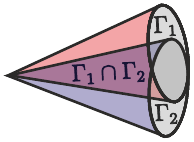}
\caption{Pictorial illustration of the NS entropic cone.
}
\label{fig:NScone}
\end{figure}

On the other hand, NS probabilities are defined as those where the outcomes of a part do not depend on the measurements performed by another distant part, i.e.,
such that $p(a_x)=\sum_{b_y}p(a_x,b_y)=\sum_{b_{y^{\prime}}}p(a_x,b_{y^{\prime}})$ (similarly for $b$ and for any number of parties). NS correlations are then defined by the above linear constraints (NS conditions) together with the nonnegativity condition $p\geq 0$; i.e., they are classical probability distributions whenever restricted to $p(a_x,b_y)$, with consistent marginals. Geometrically, they can be seen as the intersection of the simplex polytopes \cite{boyd_convex_2009} defining each of the probabilities $p(a_x,b_y)$ and thus overlapping over the marginals $p(a_x)$ and $p(b_y)$. We can then naturally define NS entropic cone, for a marginal scenario  $\mathcal{M}=\left\{ S_1,\dots, S_{\vert \mathcal{M} \vert} \right\}$, as the intersection $\Gamma_{\mathrm{NS}}=\Gamma_i \cap \dots \cap \Gamma_{\vert \mathcal{M} \vert}$, where $\Gamma_i$ is the entropy cone associated with $S_i$ (see \figref{fig:NScone}). For instance, in the bipartite scenario, the NS cone is given by the intersection of $2m$ cones corresponding to the subsets of variables appearing in the marginal scenario  $\mathcal{M}_{\mathrm{Bell}}=\left\{\left\{A_x,B_y \right\}\right\}_{x,y=1,\ldots,m}$ and respecting the basic constraints given by $H(A_x \vert B_y) \geq 0$, $H(B_y \vert A_x) \geq 0$ and $I(A_x:B_y) \geq 0$. This intersection can be understood as follows: since each $S_i$ contains a restricted set of variables, we embed each $\Gamma_i$ in a bigger space where the variables not in $S_i$ are unconstrained.

In the following, we apply the above framework to analyze from an entropic perspective a broad range of scenarios. Notice that for $n\leq 3$ variables, the entropy cone corresponds to the Shannon cone, i.e., $\Gamma_{\rm E}^n = \Gamma_{\rm Sh}^n$ \cite{Yeung2008}; hence, all results for the bipartite and tripartite cases lead to the exact description of the NS cones \cite{fnote5}. Further discussions and technical details can be found in \cite{SM}, including the derivation of all Bell inequalities that, nicely, can be proven by simple sums of Shannon inequalities.

\textit{Bipartite and tripartite scenarios.---}
We start with the simplest Bell scenario as above, for $m=2$. In contrast to the probabilistic case, entropic correlations are concisely defined for an arbitrary number of measurement outcomes, highlighting another advantage of the entropic approach. For dichotomic observables ($a_x,b_y=0,1$), the only nontrivial probabilistic Bell inequality is the Clauser-Horne-Shimony-Holt (CHSH) inequality \cite{Clauser1969}
\begin{equation}
\label{CHSH}
S=\mean{A_0B_0}+\mean{A_0B_1}+\mean{A_1B_0}-\mean{A_1B_1}-2 \leq 0,
\end{equation}
where $\mean{A_xB_y}=\sum (-1)^{a_x+b_y}p(a_x,b_y)$ stand for the expectation value. Its entropic version \cite{Braunstein1988,Chaves2012},
\begin{equation}
\label{ECHSH}
S_{\mathrm{E}}=I_{A_0:B_0} +I_{A_0:B_1} +I_{A_1:B_0}-I_{A_1:B_1} -H_{A_0}-H_{B_0} \leq 0,
\end{equation}
is valid for any number of outcomes, where $I_{A_x:B_y}=H_{A_x}+H_{B_y}-H_{A_xB_y}$ represents the mutual information.

Both inequalities are maximally violated by an extremal point or ray characterizing the NS correlations. Eq.~\eqref{CHSH} is maximally violated by the Popescu-Rohrlich (PR)-box $p_{\mathrm{PR}}(a_x,b_y)=(1/2)\delta_{a\oplus b,xy}$. However, $p_{\mathrm{PR}}$ is entropically equivalent to the classical correlation $p_{\mathrm{C}}(a_x,b_y)=(1/2)\delta_{a\oplus b,0}$ and thus cannot violate \eqref{ECHSH}. On the other hand, Eq.~\eqref{ECHSH} is maximally violated by $H(A_x,B_y)=\log_2(d)(1+xy)$, with marginals $H(A_x)=H(B_y)=\log_2(d)$ and $d$ the number of outcomes. Thus, this correlation can be interpreted as the entropic counterpart of a PR-box. For $d=2$, these entropies are obtained as an equal mixture of $p_{\mathrm{PR}}$ and $p_{\mathrm{C}}$. The mixing with $p_{\mathrm{C}}$ is exactly the method proposed in \cite{Chaves2013a} to turn entropic inequalities into necessary and sufficient conditions for nonlocality detection. It is thus appealing that the NS entropic cone naturally retrieves this sort of correlations.

Another important result of our approach is the derivation of the first entropic monogamy relation for Bell inequalities.
The monogamy of Bell inequalities violations is a general feature of NS theories \cite{Masanes2006}, and it can be understood by the following example. For a tripartite distribution $p(a_x,b_y,c_z)$ with binary inputs and outputs, whenever the marginal distribution $p(a_x,b_y)$ violates the CHSH inequality necessarily $p(a_x,c_z)$ must be local. Similarly, from the definition of NS entropic cone, we can prove that
\begin{equation}
\label{monogamy}
S^{AB}_{\mathrm{E}}+S^{AC}_{\mathrm{E}}\leq 0,
\end{equation}
meaning that both entropic Bell inequalities, between Alice-Bob and Alice-Charlie, cannot be violated at the same time, with the notable difference that this monogamy inequality is valid for any number of outcomes.

The similarities between the probabilistic and entropic approaches, which may suggest a deeper geometric connection \cite{Chaves2014}, already disappear in the tripartite scenario. For the case of three parties and two settings, the probabilistic NS correlations for dichotomic measurements consist of $46$ different classes of extremal points, with $45$ of them nonlocal \cite{pironio2011extremal}. In turn, the entropic NS cone is characterized by $1292$ different classes of extremal rays, $1164$ of which correspond to nonlocal correlations \cite{fnote4}.
As it turns out, already at the tripartite case we obtain a much more complex structure than the one we could naively presume from the probabilistic description.

\textit{Genuine tripartite entropic nonlocality.---}
In analogy to entanglement \cite{horodecki2009quantum}, when moving beyond the bipartite
case, different classes of nonlocality arise. With three parties, one can introduce the
notion of genuine tripartite nonlocality, that is, a stronger form of nonlocality that
cannot be reproduced even if any two of the parties are allowed to share some nonlocal
resources \cite{Svetlichny1987,Gallego2012,Pironio2013}.
We focus our attention to nonsignaling resources (e.g., a PR-box) and two possible
measurements per party, extensions to more measurements and parties are straightforward. For a given bipartition, say $A|BC$, a hybrid local-nonsignaling (L$|$NS)
model is equivalent to the existence of probability distributions $p(a_0,a_1,b_j,c_k)$, with
consistent marginal $p(a_0,a_1,b_j), p(a_0,a_1,c_k)$, i.e., Alice has local correlations and
Bob and Charlie share nonsignaling correlations. Genuine tripartite nonlocal (GTNL)
correlations correspond to marginals $p(a_i,b_j,c_k)$ that cannot be explained as a convex
combination of models of the type $A|BC$, $B|AC$, and $C|AB$.

Analogously to the NS case, an entropic $A|BC$ model corresponds to the joint entropies
$H(A_0,A_1,B_j,C_k)$, $j,k=0,1$, and all its marginals, and similarly for $B|AC$, and
$C|AB$.  We can then define the L$|$NS entropic correlations via the cone
$\Gamma_{\mathrm{L}|\mathrm{NS}}$, constructed as the convex hull (i.e., set of convex
combinations) of the entropic cones for each of the models $A|BC$, $B|AC$, and $C|AB$. In
turn, GTNL entropic correlations are those lying outside $\Gamma_{\mathrm{L}|\mathrm{NS}}$.

\begin{figure}[ht]
\vspace{0.6cm}
\center
\includegraphics[width=0.8\columnwidth]{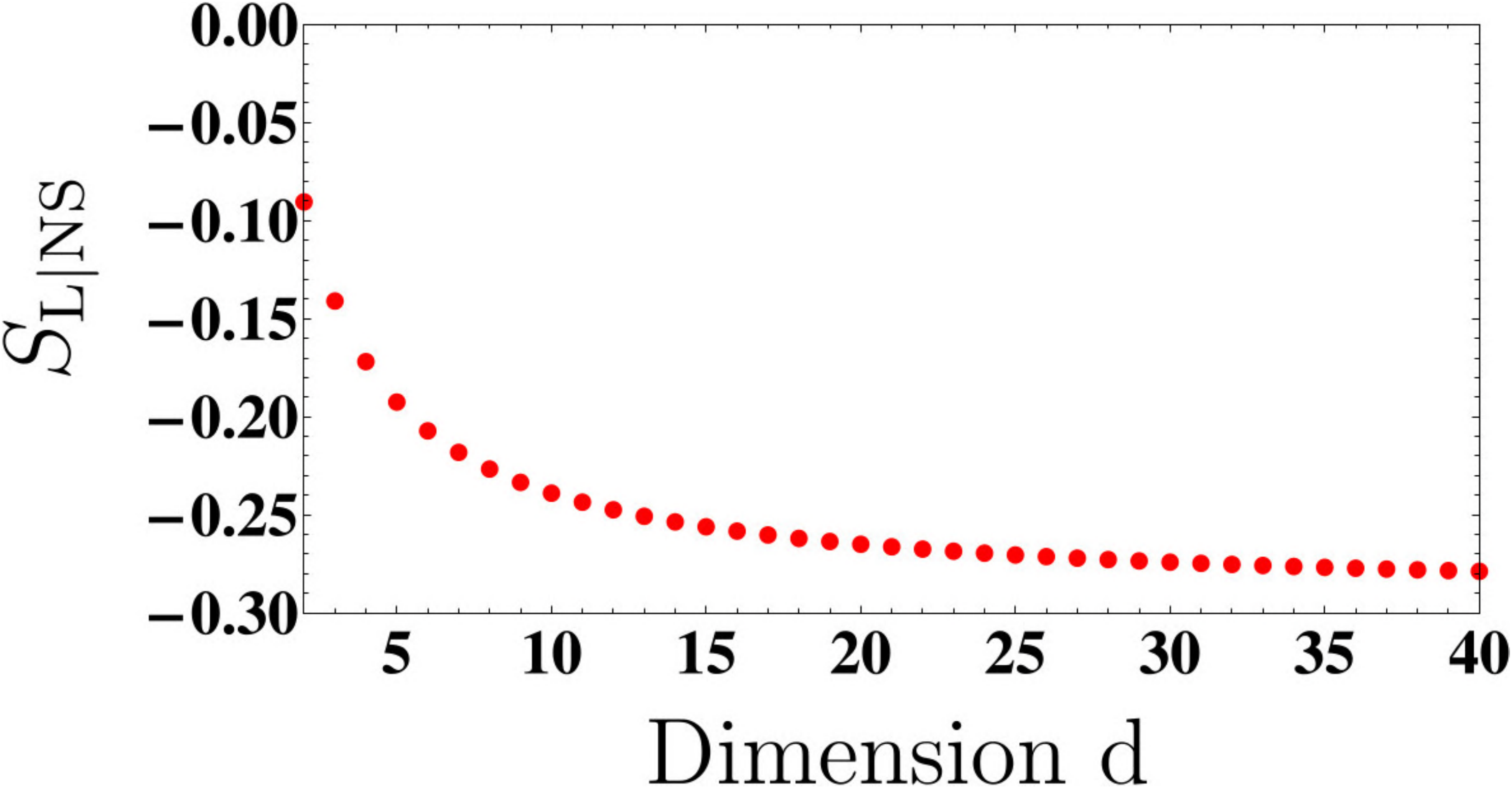}
\caption{The violation of inequality $S_{\mathrm{L}|\mathrm{NS}} \geq 0$ using GHZ states. The points stand for the violation obtained via numerical optimization.
}
\label{fig:GTNL_GHZ}
\end{figure}

From the $1164$ different classes of extremal nonlocal rays defining the tripartite
scenario, $932$ correspond to GTNL correlations. One of these rays correspond to the
distribution $(1/2)(p_{\mathrm{XYZ}}+p_{\mathrm{C}})$, that is, the mixing of
$p_{\mathrm{XYZ}}(a,b,c \vert x,y,z)=(1/4)\delta_{a\oplus b \oplus c,xyz}$
\cite{Barrett2005b} with classical correlations
$p_{\mathrm{C}}(a,b,c \vert x,y,z)=(1/4)\delta_{a\oplus b \oplus c,0}$. The GTNL character
of this correlation can be witnessed by the violation of the following entropic inequality
valid for any L$|$NS correlation with arbitrary number of outcomes:
\begin{eqnarray}
\nonumber
\label{EGTNL}
& & S_{\mathrm{L}|\mathrm{NS}} = H_{A_1B_1C_0} + H_{A_1B_0C_0} + H_{A_1B_0C_1} + H_{A_0B_1C_0}  \\
& &- H_{A_1B_1C_1} -H_{A_1B_0}- H_{A_1C_0}- H_{A_0C_1} - H_{B_1C_0} \geq 0.
\end{eqnarray}
Furthermore, inequality \eqref{EGTNL} can also be used to witness the GTNL in quantum
states, for instance using $d$-dimensional Greenberger-Horne-Zeilinger (GHZ) states
$\ket{\mathrm{GHZ}}=(1/\sqrt{d})\sum_{j=0}^{d-1} \ket{jjj}$ and projective measurements \cite{Collins2002}. Results are plotted in Fig.~\ref{fig:GTNL_GHZ} up to $d=40$.

\textit{Activating entropic nonlocality in networks.---}
The tripartite scenario permits also another possibility: that the correlations between the
parties are mediated by independent sources. The paradigmatic example is the entanglement
swapping experiment \cite{Zukowski1993}. Two independent pairs of entangled particles are
distributed among three spatially separated parties: Bob receives one particle of each pair,
and Alice and Charlie the remaining two. By jointly measuring his particles, Bob can
generate (upon conditioning on outcomes) entanglement and nonlocal correlations between the
two remaining particles, even though the latter have never interacted. A probabilistic and local
realistic description of this experiment involves two independent hidden variables, the so
called bilocality assumption \cite{Branciard2010,Tavakoli2014,Chaves2015b,Chaves2016,Rosset2016}, implying the independence relation
$p(a,c)=p(a)p(c)$, i.e., no correlations between Alice and Charlie. The local and NS
correlations in the bilocality scenario are defined by infinitely many extremal points and
is extremely challenging to characterize \cite{Chaves2016}.

The advantages of the entropic description are here apparent: independence
constraints are encoded in simple linear relations, e.g.,
${p(a,c)=p(a)p(c) \rightarrow I(A:C)=0}$. Geometrically,
a set of extra linear constraints
$L\h=0$, as the one above, corresponds to the intersection of the (polyhedral) convex cone
(e.g.,$\Gamma_{\mathrm{Bell}}$ or $\Gamma_{\mathrm{NS}}$)  with a linear subspace, which is
still a (polyhedral) convex cone \cite{fnote2}.

For the case of two settings per party, we have fully characterized the set of NS bilocal correlations: we found $329$ different classes of extremal rays, of which $314$ are nonlocal. Out of these, $40$ are genuinely nonbilocal, i.e., the correlations admit a LHV model but not a bilocal LHV model. A particularly interesting extremal correlation is the following: $H(A_0,B,C)=H(A_1,B,C)=H(A_1,B)=H(A_0,C)=H(A_1,C)=H(B,C)=2$ and $H(A_0,B)=H(A_0)=H(A_0)=H(B)=H(C)=1$. It can be understood as the case where Bob and Charlie always measure the same observable (no measurement choice) while Alice still can perform two different measurements. Clearly, since only one of the parties has measurement choices, all correlations arising in this scenario are compatible with a LHV model. However, this correlation is not bilocal, as it can be witnessed by the violation of the entropic inequality valid for any bilocal decomposition:
\begin{equation}
\label{bilocal_ineq1}
H(A_0,C) \leq H(A_0,B)+ H(C \vert A_1,B).
\end{equation}
The entropic correlation above arises from a probability distribution $p(a,b,c \vert x)=(1/4)\delta_{a\oplus b,xc}$, obtained when Alice and Bob share a PR-box $p_{\mathrm{PR}}$
while Bob and Charlie share a classical correlated distribution $p_{\mathrm{C}}$. To that
aim, Bob assigns $b$ to the output of his share of the PR-box that takes as input the bit
that is classically correlated with the output $c$ of Charlie. This result illustrates two
novel aspects of the bilocality scenario. First, we see that the nonlocality of the PR-box,
which in CHSH scenario is entropically equivalent to a classical correlation, can be
activated by employing it in a network. Even more remarkable is the fact that the emergence
of nonlocal correlations only requires one out of the three parties to have access to
measurement choices. This is similar to what has been observed in
\cite{Branciard2012,Fritz2012}, where it has been argued that since the role of Charlie can be
interpreted as defining measurement choices for Bob, this scenario can be mapped to the CHSH
one. In our case, however, we do not need to hinge on this mapping, since we violate a new
sort of entropic Bell inequality. Thus, as opposed to Refs. \cite{Branciard2012,Fritz2012},
our result does not rely on Bell's theorem.

Another genuinely nonbilocal extremal entropic ray is associated with the probability $p(a,b,c\vert x,y,z)=(1/8)(\delta_{a\oplus b \oplus c,xyz\oplus xy \oplus xz \oplus yz \oplus z \oplus 1}+\delta_{a \oplus b \oplus c,0})$. Its nonbilocality can be witnessed via the violation of
\begin{eqnarray}
\label{bilocal_ineq2}
S_{\mathrm{BL}}= & & -H_{A_0B_0C_0}+H_{A_1B_1C_0}  \\ \nonumber
& & +H_{A_0B_0C_1}+H_{A_1B_1C_1}-H_{A_1,C_1}-H_{A_1,B_1} \geq 0,
\end{eqnarray}
specifically, with value $S_{\mathrm{BL}}=-1$. As opposed to other known inequalities \cite{Branciard2010,Branciard2012,Chaves2016,Rosset2016}, Eq.~\eqref{bilocal_ineq2} includes marginal terms, and it is valid for an arbitrary number of outcomes.

\textit{Information causality.---}
Information causality (IC) \cite{Pawlowski2009} is a principle introduced to explain the limitation of quantum correlations, i.e., Tsirelson bound \cite{Cirel1980quantum}. It can be understood as a game: Alice receives two independent random bits $X_0$ and $X_1$ and the task of Bob is to guess, at each run of the experiment, the value of one of them, having as resources some preshared correlations with Alice and some classical communication ($H(M)$ bits) sent by her. For shared quantum correlations, the following inequality holds \cite{Pawlowski2009}:
\begin{equation}
\label{IC_ineq}
I(X_0:G_0)+I(X_1:G_1) \leq H(M).
\end{equation}
where $G_{s}$ denotes Bob's guess of $X_s$.

To characterize the set of NS entropic correlations associated to IC scenario [i.e,
including post-quantum correlations violating \eqref{IC_ineq}], first notice that the mutual
information between Alice's inputs and Bob's guesses should be limited, according to the assumed causal structure, by the amount of
communication, that is, $I(X_s:G_s) \leq H(M)$, otherwise they could also communicate
superluminally \cite{popescu2014nonlocality}. Here, similarly to what has been done in
\cite{Pawlowski2009}, we consider the marginals $\mathcal{M}_{\textrm{IC}}=\left\{ \left\{ X_0,G_0 \right\},\left\{ X_1,G_1 \right\},\left\{ M \right\}  \right\}$. The NS cone
$\Gamma_{\mathrm{IC}}$ is thus given by the intersection of the Shannon cone defined by
$\mathcal{M}_{\textrm{IC}}$ with the constraints $I(X_y:G_y) \leq H(M)$ ($y=0,1$) arising
from the causal structure of the game \cite{Chaves2015a}. We found $\Gamma_{\mathrm{IC}}$ to
be characterized by $8$ extremal rays, $7$ of which respect Eq.~\eqref{IC_ineq}. The
extremal ray violating Eq.~\eqref{IC_ineq} is given by
$H(X_0)=H(X_1)=H(G_0)=H(G_1)=H(X_0,G_0)=H(X_1,G_1)=H(M)=1$. It is achieved when the parties
share a PR-box and apply the protocol used in \cite{Pawlowski2009}.
It is once more appealing that the NS cone approach naturally retrieves an entropic correlation of special importance.

\textit{Discussion.---}
Nonlocality stands nowadays as one of the cornerstones in our understanding of quantum
theory. In turn, entropy is a key concept in the foundations and applications of quantum
information science. It is thus surprising that still so little is known about their
relations and in particular what nonsignaling---another guiding principle permeating all
physics---has to say about the entropies that can be generated by the outcomes of physical
measurements.
Here, we introduced the notion of entropic nonsignaling correlations characterizing the
entropies compatible with the fact that we cannot transmit information instantaneously. To
illustrate its relevance and novelty, we have applied it to understand a broad range of
different phenomena from an entropic perspective: from monogamy relations and nonlocality activation in networks, to
genuine multipartite nonlocality.

Nonsignaling also lies at the heart of the device-independent approach to quantum information, which has lately attracted growing attention  \cite{Ekert1991,Barrett2005,colbeck2009quantum,pironio2010random,colbeck2012free,Gallego2010,Chaves2015entropy}, and we believe our results provide a new tool also for practical applications.
In addition, the entropic approach provides the natural ground to treat generalized Bell scenarios \cite{Tavakoli2014,Chaves2015b,
 Chaves2016,Rosset2016} and understand novel forms of nonlocal correlations emerging from it. Future lines of research also include monogamy relations \cite{Pawlowski2009b}, the role of non-Shannon type inequalities \cite{Yeung2008} in multipartite scenarios and possible applications in nonlocal games \cite{Brukner2004}.

 Finally, as demonstrated by information causality \cite{Pawlowski2009,Chaves2015a}, many of our current guiding principles are stated in terms of entropy. Our current framework can help to devise new entropic principles, in particular for the multipartite case \cite{gallego2012quantum}.

\begin{acknowledgments}
The authors thank Nikolai Miklin for discussions. R.C. acknowledges financial support from the Excellence Initiative of the German Federal and State Governments (Grants ZUK 43 \& 81), the FQXi Fund, the US Army Research Office under Contracts W911NF-14-1-0098 and W911NF-14-1-0133 (Quantum Characterization, Verification, and Validation), the DFG (GRO 4334 \& SPP 1798). C.B. acknowledges financial support from the EU (Marie Curie CIG 293993/ENFOQI), the FQXi Fund (Silicon Valley Community Foundation), and the DFG.
\end{acknowledgments}

\section{Appendix}

\subsection{The Shannon and Bell entropic cones}\label{app:cone}
Given a collection of $n$ discrete random variables $X_1, \dots, X_n$, we denote by $[n]=\{1, \dots, n\}$ the set of indices and $2^{[n]}$ its power set. For every $S\in 2^{[n]}$, let $X_S$ be the vector $(X_i)_{i\in S}$ and $H(S):=H(X_S)$  the associated Shannon entropy,  given by $H(X_S) :=-\sum_{x_s}p(x_s)\log_2 p(x_s)$.
We can define the vector $\h=(H(\emptyset),H(X_1), \ldots, H(X_1, \dots, X_n)) \in R_n := \mathbb{R}^{2^n}$. Not every vector $\h\in R_n$ will correspond to an entropy vector, as, e.g., entropies are nonnegative. The entropy cone is defined as the closure of the region
\begin{eqnarray*}
	\Gamma_{\rm E} := \overline{\left\{ h \in R_n \,|\, h_S = H(S) \text{ for some entropy
	} H \right\}}.
\end{eqnarray*}
$\Gamma_{\rm E}$ is known to be a convex cone but a tight and explicit description is still to be found \cite{Yeung2008}. However, an outer approximation to the entropic cone is known, the so-called Shannon cone.

The Shannon cone $\Gamma_{\mathrm{Sh}}$ is a polyhedral closed convex cone, i.e., a subset of $R_n$ defined by a finite set of linear inequalities, known as basic Shannon-type inequalities, plus a normalization constraint. The first type of inequalities are given by monotonicity conditions, for example, $H( X_{S \cup\{i\}}) \geq H(X_{S})$, stating that the uncertainty about a set of variables should always be larger than or equal to the uncertainty about any subset of it. The second type is the strong subadditivity condition given by $I(X_{i}:X_{ j} \vert X_S)= H(X_{S \cup i}) +H(X_{ S \cup j})-H( X_{S \cup\{i,j\}})-H(X_{S}) \geq 0$ which is equivalent to the positivity of the conditional mutual information. The normalization constraint imposes that $H(\emptyset)=0$.

Given $n$ variables, all the associated Shannon type inequalities (and thus the Shannon cone) are characterized by the following elemental (non-redundant) inequalities \cite{Yeung2008}
\begin{eqnarray}
	H_{}([n]\setminus\{i\}) &\leq& H_{}([n]),
	\label{shannonineqs_basic}\\
	H_{}(S) + H_{}(S\cup\{i,j\}) &\leq& H_{}(S\cup \{i\}) + H_{}(S\cup \{j\}),
	\nonumber
	\\
	H_{}(\emptyset) &=& 0,
	\nonumber
\end{eqnarray}
for all $S \subset [n] \setminus\{i,j\}$, $i \neq j$ and $i, j\in [n]$. Thus, the Shannon cone associated with $n$ variables is described by $2^{n-2}\binom{n}{2}+n$ inequalities plus one normalization constraint.

Given a marginal scenario $\mathcal{M}$, we are interested in the projection of $\Gamma_{\mathrm{Sh}}$ in the subspace of $R_n$ representing only observable terms, that is, the Bell cone $\Gamma_{\mathrm{Bell}}$ associated with the marginal scenario in question. Since we are given a description of the Shannon cone in terms of linear inequalities, to perform this projection we need to eliminate from this system of inequalities all terms corresponding to non-observables terms. In practice, this is achieved via a Fourier-Motzkin (FM) elimination \cite{Williams1986}. The final set of inequalities obtained via the FM elimination (and after eliminating over redundant inequalities) gives all the facets of the associated Bell cone, the non-trivial of which are exactly the entropic Bell inequalities \cite{Braunstein1988,Chaves2012,FritzChaves2013,Chaves2014}. By non-trivial, we denote those inequalities that are not simply basic Shannon type inequalities like \eqref{shannonineqs_basic}.

To exemplify the general framework it is useful to analyze the particular case of the CHSH scenario discussed in the main text. In this case, the existence of an underlying LHV model implies the existence of $H(A_0,A_1,B_0,B_1)$ (and all its marginals) respecting the elementary inequalities in \eqref{shannonineqs_basic}. Proceeding with the FM elimination and keeping only the observable terms $H(A_i,B_j)$, $H(A_i)$ and $H(B_j)$ (with $i,j=0,1$) we observe that the only non-trivial inequality (up to permutations) is given by
\begin{eqnarray}
\label{ECHSH_ap}
& & H(A_0,B_0)+H(A_0,B_1)+H(A_1,B_0) \\ \nonumber
& &-H(A_1,B_1)-H(A_0)-H(B_0) \geq 0.
\end{eqnarray}
This is exactly the inequality originally derived in \cite{Braunstein1988}. Replacing the bipartite entropies by mutual informations in \eqref{ECHSH_ap}, that is, using $H(A_x,B_y)=H(A_x)+H(B_y)-I(A_x:B_y)$, we obtain the form (2) discussed in the main text and that can be understood as the entropic analogue of the CHSH inequality. However, opposed to the probabilistic version of the CHSH inequality, its entropic version can be used as a nonlocality witness for an arbitrary finite number of measurement outcomes.

\subsection{A tool for the derivation of entropic Bell inequalities}
\label{subsec:tool}
The formalism outlined above provides a general framework for the derivation of entropic Bell inequalities in basically any scenario \cite{Chaves2012,FritzChaves2013,Chaves2014}. In short, we have to perform the projection --via a Fourier-Motzkin (FM) elimination -- of the Shannon cone $\Gamma_{\mathrm{Sh}}$ to the subspace of observable coordinates defining the Bell cone $\Gamma_{\mathrm{Bell}}$.
The problem with this approach is that it turns out to be computationally extremely demanding. The FM algorithm eliminate a variable $x$ from a system of inequalities by summing, after proper normalization, inequalities where $x$ appears with coefficient $+1$ and with coefficient $-1$, and keeping the rest. Eliminating over $m$ variables in a system of $N$ inequalities can lead to a number $O(N^{2^m})$ of inequalities, that is, double exponential. Since the number of initial inequalities describing $\Gamma_{\mathrm{Sh}}$ is itself exponential \cite{Yeung2008}, this leads to a triple-exponential complexity algorithm, which is limited, in practice, to very simple cases. In fact, no systematic characterization of entropic inequalities is known beyond particular instances of the bipartite case, although particular multipartite inequalities have been derived \cite{Chaves2014,raeisi2015entropic}. In particular, no entropic Bell inequality witnessing genuine multipartite nonlocality \cite{Svetlichny1987} was known to this date.

Nevertheless, checking whether a given inequality is valid for a scenario of interest is computationally much simpler. First, notice
that any valid entropic inequality must follow from a FM elimination over the system of
inequalities defining the scenario at question, for example, entropic Bell inequalities
follow from the basic Shannon type inequalities \eqref{shannonineqs_basic} plus possible additional linear constraints.
Given the entropy vector $\boldsymbol{h} \in R_n$, any linear inequality can be written
as the inner product $\langle \boldsymbol{\mathcal{I}}, \boldsymbol{h} \rangle \geq 0$, where $\boldsymbol{\mathcal{I}}$ is a vector to the inequality. Similarly, a system of inequalities, e.g., Eq.~\eqref{shannonineqs_basic}, can be represented as a matrix $M$,
such that $\Gamma := \{ \boldsymbol{h} | M\boldsymbol{h} \geq \boldsymbol{0}\}$.
To check the validity of an inequality $\langle \boldsymbol{\mathcal{I}}, \boldsymbol{h} \rangle \geq 0$ with respect to the system $M\boldsymbol{h} \geq \boldsymbol{0}$, one simply needs to solve the following (efficient) linear program \cite{Yeung2008}:
\begin{eqnarray}
\underset{\boldsymbol{h} \in \RR^{2^n}}{\minimize} & & \langle \boldsymbol{\mathcal{I}}, \boldsymbol{h} \rangle  	\label{LP} \\
\st & &  M\boldsymbol{h} \geq \boldsymbol{0}. \nonumber
\end{eqnarray}
If the minimum of $\langle \boldsymbol{\mathcal{I}}, \boldsymbol{h} \rangle$ is larger or equal to zero, then the inequality is valid.

Moreover, one can extend this result to inequalities for the projected cone, even without performing the corresponding FM elimination on the system $M\boldsymbol{h} \geq \boldsymbol{0}$.  More precisely, given
${\Gamma := \{\ \boldsymbol{h}\ |\ M\boldsymbol{h}\geq 0\ \}}$ and
${\Pi_k(\Gamma) = \{\ \boldsymbol{h}\ |\ \tilde{M}\boldsymbol{h}\geq 0,\ (\boldsymbol{h})_l = 0 \text{ for } l> k\ \}}$, where $\Pi_k$
is the projection on the first $k$ coordinates, we prove that the inequality $\langle \boldsymbol{\mathcal{I}}, \boldsymbol{h} \rangle\geq 0$, with
$(\boldsymbol{\mathcal{I}})_l = 0$ for $l>k$ is valid for the polyhedral cone
$\Pi_k(\Gamma)$ if and only if it is valid for the polyhedral cone $\Gamma$.
If there exists $\boldsymbol{p} \in \Gamma$
such that $\langle \boldsymbol{\mathcal{I}}, \boldsymbol{p} \rangle < 0$, then
$\Pi_k( \boldsymbol{p})\in  \Pi_k(\Gamma)$ and $\langle \boldsymbol{\mathcal{I}}, \Pi_k( \boldsymbol{p}) \rangle = \langle \boldsymbol{\mathcal{I}}, \boldsymbol{p} \rangle$. Vice
versa, given $\tilde{\boldsymbol{p}}\in \Pi_k(\Gamma)$ such that
$\langle \boldsymbol{\mathcal{I}}, \tilde{\boldsymbol{p}} \rangle<0$, there exists
$\boldsymbol{p} \in \Gamma$ such that $\Pi_k(\boldsymbol{p})=\tilde{\boldsymbol{p}}$ and
$\langle \boldsymbol{\mathcal{I}},  \boldsymbol{p} \rangle = \langle \boldsymbol{\mathcal{I}}, \tilde{\boldsymbol{p}} \rangle$.

Similarly, it is again a linear program (in fact, a feasibility problem) to check if a given observed entropic correlation $\tilde{\h}_{\mathrm{obs}}$ is local or nonlocal. It is simply given by
\begin{eqnarray}
\underset{\boldsymbol{h} \in \RR^{2^n}}{\minimize} & & \langle \boldsymbol{\mathcal{I}}, \boldsymbol{h} \rangle  	\label{LP2} \\
\st & &  M\boldsymbol{h} \geq \boldsymbol{0}, \\ \nonumber
& & \h_{\mathrm{obs}}=\tilde{\h}_{\mathrm{obs}}.
\end{eqnarray}
where $\boldsymbol{\mathcal{I}}$ in this case can be any linear objective function. We see that in this case, we not only impose the constraints $ M\boldsymbol{h} \geq \boldsymbol{0}$ but also impose that some of the coordinates of the vector $\boldsymbol{h}$ (those given by $\h_{\mathrm{obs}}$) should correspond to the observable quantities $\tilde{\h}_{\mathrm{obs}}$. In case the linear program above has no solution (non-feasible), the correlations $\tilde{\h}_{\mathrm{obs}}$ at question are thus nonlocal.

These linear programs together with the characterization of the entropic NS cone provide novel tools in the derivation of entropic Bell inequalities, as explained below.

A nonlocal extremal correlation given by $\tilde{\h}^{\mathrm{NS}}_{\mathrm{obs}}$ will
imply a non-feasible LP of the form \eqref{LP2}. However, not all the constraints encoded in
the matrix $M$ will be necessary for witnessing this non-feasibility: many of the
inequalities in $M$ can be eliminated until we reach a (not unique) minimum set of
inequalities that will still lead to a non-feasible LP. Given this minimum set we can then
perform a FM elimination that thus will lead to an inequality that is violated by the given
correlation $\tilde{\h}^{\mathrm{NS}}_{\mathrm{obs}}$. Notice that in general the obtained
inequalities are not necessarily going to be facets of the marginal cone of interest.
However, we can obtain tight inequalities by adding noise to the extremal correlations (for
instance, white noise). By doing that we guarantee that we are deriving inequalities probing
the nonlocal character of the given correlation until the noise is strong enough to make it
enter in the marginal cone of interest (and thus become local). Notice that a similar
procedure can be applied to obtain the minimum set of inequalities required to prove the
validity of a given inequality bounding the marginal cone of interest.

\subsection{Entropic NS cones in a variety of scenarios}
For the entropic NS cone description, we use the elemental Shannon type inequalities
\eqref{shannonineqs_basic} for each subset of mutually compatible variables defined by a
given marginal scenario. In the sections below, we describe in details the entropic NS cone
in the bipartite and tripartite scenario, and the hybrid local-nonsignaling models. Notice
that for $n\leq 3$, the Shannon cone correspond with the true entropy cone, so our bounds
are tight. For each of the scenarios we consider, we have catalogued all
the inequivalent classes of extremal rays and characterized whether they correspond to local
or nonlocal correlations.

\subsubsection{Bipartite}
In a bipartite Bell scenario, two parties Alice and Bob can measure $m$ possible different observables. Thus, the bipartite entropic NS cone $\Gamma^{2}_{\mathrm{NS}}$  is defined by the elemental inequalities for each subset of mutually compatible observables $A_x,B_y$, that is, it is simply described in terms of the elemental inequalities $I(A_x:B_y) \geq 0$, $H(A_x,B_y) \geq H(A_x)$ and $H(A_x,B_y) \geq H(B_y)$ for $x,y=0,\dots,m-1$.

As discussed in the main text, for the case $m=2$ (corresponding to the CHSH scenario) $\Gamma^{2}_{\mathrm{NS}}$ is characterized by $5$ different classes of extremal NS rays, $4$ of which correspond to local correlations and only $1$ correspond to nonlocal correlations. In Table \eqref{tab:bipartite} we list all the different classes (a class is defined by the inequalities that are equivalent up to the permutation of parties and/or observables).

\begin{table*}
\begin{tabular}{|c| c c| c c| c c c c|} \hline
\multicolumn{9}{|c|}{Extremal ray}\\
\hline
      \multicolumn{1}{|c|}{\textbf{\#}}
    & \multicolumn{2}{c|}{\scriptsize$H(A_x)$}
    & \multicolumn{2}{c|}{\scriptsize$H(B_y)$}
    & \multicolumn{4}{c|}{\scriptsize$H(A_xB_y)$}\\
\small{$x$ / $y$ / $xy$}
&\small{0}&\small{1}
&\small{0}&\small{1}
&\small{00}&\small{01}&\small{10}&\small{11}
\\
\hline
\textbf{1}
&\small{1}&\small{0}
&\small{0}&\small{0}
&\small{1}&\small{1}&\small{0}&\small{0}
\\
\hline
\textbf{2}
&\small{1}&\small{0}
&\small{1}&\small{0}
&\small{1}&\small{1}&\small{1}&\small{1}
\\
\hline
\textbf{3}
&\small{1}&\small{1}
&\small{1}&\small{0}
&\small{1}&\small{1}&\small{1}&\small{1}
\\
\hline
\textbf{4}
&\small{1}&\small{1}
&\small{1}&\small{1}
&\small{1}&\small{1}&\small{1}&\small{1}
\\
\hline
\textbf{5}
&\small{1}&\small{1}
&\small{1}&\small{1}
&\small{1}&\small{1}&\small{1}&\small{2}
\\
\hline
\end{tabular}
\caption{Inequivalent classes of extremal rays in the bipartite scenario where both parties measure two possible observables. Extremal rays $1$-$4$ correspond to local entropies while ray $5$ correspond to the entropic analog of the PR-box \cite{Popescu1994}. All rays are defined up to a positive constant multiplicative factor, that is, the rays are described by a vector in some real space $v\in\mathbbm{R}^n$ such that $\{ \lambda v \,|\, \lambda \geq 0\}$}. \label{tab:bipartite}
\end{table*}

The class $\# 1$ of extremal rays corresponds to the case where one of the variables has maximal entropy, e.g, $H(A_0)= \log_2 d$ (where $d$ is the number of outcomes), while all other variables have null entropy (that is, they represent probability distributions with deterministic outcomes). Class $\# 2$ represents perfect (anti)correlations between one observable of Alice and one of Bob, e.g., $H(A_0)=H(B_0)=I(A_0:B_0)=\log_2 d$ while all other entropies are null. Class $\# 3$ represents perfect (anti)correlations between the two observables of Alice with one observable of Bob, e.g, $H(A_x)=H(B_0)=I(A_x:B_0)=\log_2 d$, while the other observable of Bob has null entropy. Class $\# 4$ represents perfect (anti)correlations between all the observables of Alice and Bob, that is, $H(A_x)=H(B_y)=I(A_x:B_y)=\log_2 d$. All these rays clearly correspond to local correlations.

In turn, the class $\# 5$ corresponds to entropic nonlocal correlations. For $d=2$, it can be understood as the entropic version of the paradigmatic PR-box \cite{Popescu1994}
\begin{equation}
\label{pr}
p_{\mathrm{PR}}\left(  a,b | x,y \right)  =\left\{
\begin{array}{ll}
1/2 & \text{, } a\oplus b =xy\\
0 & \text{, otherwise}%
\end{array}
\right. .
\end{equation}
First, notice that $p_{\mathrm{PR}}$ is entropically equivalent to the purely classical correlations \cite{Chaves2013a}
\begin{equation}
\label{pc}
p_{\mathrm{PC}}\left(  a,b | x,y \right)  =\left\{
\begin{array}{ll}
1/2 & \text{, }  a \oplus b =0\\
0 & \text{, otherwise}%
\end{array}
\right. .
\end{equation}
That is, both \eqref{pr} and \eqref{pc} correspond to the class $\# 4$ of extremal rays. The reason for this equivalence is due to the fact that entropies are unable to distinguish between correlations and anti-correlations \cite{Chaves2013a}. Interestingly, the class $\# 5$ correspond to a probability distribution obtained by mixing with equal weights the distributions \eqref{pr} and \eqref{pc}, that is, $p_{\mathrm{class} \# 5}=(1/2)(p_{\mathrm{PR}}+p_{\mathrm{PC}})$ corresponding to three perfect (anti)correlated pairs of variables $I(A_0:B_0)=I(A_0:B_1)=I(A_1:B_0)=1$ and one uncorrelated pair $I(A_1:B_1)=0$. The class $\# 5$ violates the entropic version of the CHSH inequality \eqref{ECHSH_ap} up to the algebraic maximum, achieving $S_{\mathrm{E}}=\log_2 d$.

We have performed the same analysis for the case $m=3$ and obtained $20$ different classes of extremal rays, $7$ of which correspond local and $13$ to nonlocal correlations. The different classes of extremal rays are listed in Table \eqref{tab:bipartite3}. The interpretation of the local rays is identical to the one we have detailed above to the case of $2$ measurement settings. Regarding the nonlocal rays, we focus attention to the classes $\# 8$ an $\# 19$. Class $\# 8$ basically correspond to the entropic version of the PR-box discussed above, since  $H(A_2)=H(B_2)=0$. In turn, class $\# 19$ can be understood as the NS correlation that maximally violates the entropic version of the Collins-Gisin inequality \cite{Collins2004} given by \cite{Chaves2014}
\begin{eqnarray}
& & S_{3}= I(A_0:B_2)-I(A_0:B_1)+I(A_1:B_1)\\ \nonumber
& & -I(A_1:B_0)+I(A_1:B_2)+I(A_2:B_2)+I(A_2:B_1) \\ \nonumber
& & +I(A_2:B_0)-H(A_2)-2H(B_2)-H(B_1)  \leq 0,
\end{eqnarray}
reaching $S_3=2\log_2 d$.

\begin{table*}
\begin{tabular}{|c| c c c| c c c| c c c c c c c c c|} \hline
\multicolumn{16}{|c|}{Extremal rays}\\
\hline
      \multicolumn{1}{|c|}{\textbf{\#}}
    & \multicolumn{3}{c|}{\scriptsize$H(A_x)$}
    & \multicolumn{3}{c|}{\scriptsize$H(B_y)$}
    & \multicolumn{9}{c|}{\scriptsize$H(A_xB_y)$}\\
\small{$x$ / $y$ / $xy$}
&\small{0}&\small{1}&\small{2}
&\small{0}&\small{1}&\small{2}
&\small{00}&\small{01}&\small{02}&\small{10}&\small{11}&\small{12}&\small{20}&\small{21}&\small{22}
\\
\hline
\textbf{1}
&\small{1}&\small{0}&\small{0}
&\small{0}&\small{0}&\small{0}
&\small{1}&\small{1}&\small{1}&\small{0}&\small{0}&\small{0}&\small{0}&\small{0}&\small{0}
\\
\hline
\textbf{2}
&\small{1}&\small{0}&\small{0}
&\small{1}&\small{0}&\small{0}
&\small{1}&\small{1}&\small{1}&\small{1}&\small{0}&\small{0}&\small{1}&\small{0}&\small{0}
\\
\hline
\textbf{3}
&\small{1}&\small{1}&\small{0}
&\small{1}&\small{0}&\small{0}
&\small{1}&\small{1}&\small{1}&\small{1}&\small{1}&\small{1}&\small{1}&\small{0}&\small{0}
\\
\hline
\textbf{4}
&\small{1}&\small{1}&\small{0}
&\small{1}&\small{1}&\small{0}
&\small{1}&\small{1}&\small{1}&\small{1}&\small{1}&\small{1}&\small{1}&\small{1}&\small{0}
\\
\hline
\textbf{5}
&\small{1}&\small{1}&\small{1}
&\small{1}&\small{0}&\small{0}
&\small{1}&\small{1}&\small{1}&\small{1}&\small{1}&\small{1}&\small{1}&\small{0}&\small{1}
\\
\hline
\textbf{6}
&\small{1}&\small{1}&\small{1}
&\small{1}&\small{1}&\small{0}
&\small{1}&\small{1}&\small{1}&\small{1}&\small{1}&\small{1}&\small{1}&\small{1}&\small{1}
\\
\hline
\textbf{7}
&\small{1}&\small{1}&\small{1}
&\small{1}&\small{1}&\small{1}
&\small{1}&\small{1}&\small{1}&\small{1}&\small{1}&\small{1}&\small{1}&\small{1}&\small{1}
\\
\hline
\textbf{8}
&\small{1}&\small{1}&\small{0}
&\small{1}&\small{1}&\small{0}
&\small{1}&\small{1}&\small{1}&\small{1}&\small{2}&\small{1}&\small{1}&\small{1}&\small{0}
\\
\hline
\textbf{9}
&\small{1}&\small{1}&\small{1}
&\small{1}&\small{1}&\small{1}
&\small{2}&\small{2}&\small{1}&\small{2}&\small{2}&\small{1}&\small{1}&\small{1}&\small{1}
\\
\hline
\textbf{10}
&\small{1}&\small{1}&\small{1}
&\small{1}&\small{1}&\small{1}
&\small{2}&\small{1}&\small{2}&\small{2}&\small{2}&\small{1}&\small{1}&\small{1}&\small{1}
\\
\hline
\textbf{11}
&\small{1}&\small{1}&\small{1}
&\small{1}&\small{1}&\small{1}
&\small{2}&\small{1}&\small{2}&\small{1}&\small{2}&\small{1}&\small{2}&\small{1}&\small{1}
\\
\hline
\textbf{12}
&\small{1}&\small{1}&\small{1}
&\small{1}&\small{1}&\small{0}
&\small{2}&\small{1}&\small{1}&\small{2}&\small{1}&\small{1}&\small{1}&\small{1}&\small{1}
\\
\hline
\textbf{13}
&\small{1}&\small{1}&\small{1}
&\small{1}&\small{1}&\small{0}
&\small{1}&\small{2}&\small{1}&\small{2}&\small{1}&\small{1}&\small{1}&\small{1}&\small{1}
\\
\hline
\textbf{14}
&\small{1}&\small{1}&\small{1}
&\small{1}&\small{1}&\small{1}
&\small{2}&\small{2}&\small{1}&\small{2}&\small{1}&\small{1}&\small{1}&\small{1}&\small{1}
\\
\hline
\textbf{15}
&\small{1}&\small{1}&\small{1}
&\small{1}&\small{1}&\small{1}
&\small{1}&\small{2}&\small{2}&\small{2}&\small{1}&\small{1}&\small{1}&\small{1}&\small{1}
\\
\hline
\textbf{16}
&\small{1}&\small{1}&\small{1}
&\small{1}&\small{1}&\small{1}
&\small{1}&\small{1}&\small{2}&\small{1}&\small{2}&\small{1}&\small{2}&\small{1}&\small{1}
\\
\hline
\textbf{17}
&\small{1}&\small{1}&\small{1}
&\small{1}&\small{1}&\small{0}
&\small{2}&\small{1}&\small{1}&\small{1}&\small{1}&\small{1}&\small{1}&\small{1}&\small{1}
\\
\hline
\textbf{18}
&\small{1}&\small{1}&\small{1}
&\small{1}&\small{1}&\small{1}
&\small{2}&\small{2}&\small{1}&\small{1}&\small{1}&\small{1}&\small{1}&\small{1}&\small{1}
\\
\hline
\textbf{19}
&\small{1}&\small{1}&\small{1}
&\small{1}&\small{1}&\small{1}
&\small{1}&\small{2}&\small{1}&\small{2}&\small{1}&\small{1}&\small{1}&\small{1}&\small{1}
\\
\hline
\textbf{20}
&\small{1}&\small{1}&\small{1}
&\small{1}&\small{1}&\small{1}
&\small{2}&\small{1}&\small{1}&\small{1}&\small{1}&\small{1}&\small{1}&\small{1}&\small{1}
\\
\hline
\end{tabular}
\caption{Inequivalent classes of extremal rays in the bipartite scenario where both parties measure three possible observables. Extremal rays $1$-$7$ correspond to local entropies while rays $8$-$20$ correspond to nonlocal correlations.} \label{tab:bipartite3}
\end{table*}

\subsubsection{Tripartite}
In a tripartite Bell scenario, three parties Alice, Bob and Charlie perform different measurements in their shares of a joint state. Thus, the tripartite entropic NS cone $\Gamma^{3}_{\mathrm{NS}}$  is defined by the elemental inequalities for each subset of mutually compatible observables $A_x,B_y,C_z$ (with $x,y,z=0,\dots,m-1$).

We have obtained the full characterization of $\Gamma^{3}_{\mathrm{NS}}$ in terms of extremal rays for $m=2$. There are $1292$ different inequivalent classes of extremal rays, $128$ of which are local and $1164$ are nonlocal. Furthermore, as discussed in details below, in the tripartite case we can introduce the notion of genuine tripartite nonlocal (GTNL) correlations. From the $1164$ different classes of extremal nonlocal rays, $932$ of them are GTNL.

There are thus $232$ rays corresponding to nonlocal correlations but displaying no GTNL. We focus our attention to a particular class of these rays, given by $H(A_x)=H(B_y)=H(C_z)=1$, $H(A_x,B_y)=H(A_x,C_z)=H(B_y,C_z)=2$ and $H(A_0,B_0,C_0)=H(A_1,B_1,C_0)=H(A_1,B_0,C_1)=H(A_1,B_1,C_1)=2$ and $H(A_1,B_0,C_0)=H(A_0,B_1,C_0)=H(A_0,B_0,C_1)=H(A_0,B_1,C_1)=3$. It can be obtained by the mixing of the nonlocal correlation
\begin{equation}
\label{NLtri}
p\left(  a,b,c | x,y,z \right)  =\left\{
\begin{array}{ll}
1/2 & \text{, } a\oplus b \oplus c =yz\oplus x \oplus y \oplus z \\
0 & \text{, otherwise}%
\end{array}
\right. ,
\end{equation}
with the classical correlation
\begin{equation}
\label{pctri}
p_{\mathrm{PC}}\left(  a,b,c | x,y,z \right)  =\left\{
\begin{array}{ll}
1/2 & \text{, }  a \oplus b \oplus c =0\\
0 & \text{, otherwise}%
\end{array}
\right..
\end{equation}

The nonlocal character of this distribution can be witnessed via the following tripartite entropic inequality (obtained via the approach described in Sec. \ref{subsec:tool})
\begin{eqnarray}
M_3= & & H_{A_0B_1C_1} - H_{A_1B_1C_1} - H_{A_1B_1C_0} - H_{A_1B_0C_1} \\ \nonumber
& &-H_{A_0B_0C_0}+H_{B_0C_0} + H_{A_1C_1} +H_{A_1B_1} \leq 0,
\end{eqnarray}
since the correlations above imply that $M_3=1$ thus violating the inequality. To prove this inequality analytically it is sufficient to consider the chain rule for entropies, implying that
\begin{eqnarray}
H_{A_1B_1C_1A_0B_0C_0}=& &H_{A_0\vert B_1C_1A_1B_0C_0}+  H_{B_0 \vert C_1A_1B_1C_0}  \\ \nonumber
& & + H_{C_1 \vert A_1B_1C_0}  +H_{A_1 \vert B_1C_0} +H_{B_1\vert C_0} +H_{C_0}.
\end{eqnarray}
Using the basic inequalities saying that $H(A) \leq H(A,B) $ and $H(A\vert B,C) \leq H(A\vert B)$, we can turn the chain rule decomposition above in the inequality
\begin{eqnarray}
H_{A_0B_1C_1}=& &H_{A_0\vert B_0C_0}+  H_{B_0 \vert C_1A_1} + H_{C_1 \vert A_1B_1} \\ \nonumber
& &  +H_{A_1 \vert B_1C_0} +H_{B_1\vert C_0} +H_{C_0},
\end{eqnarray}
that can be rewrite exactly as $M_3 \leq 0$ if we use that $H(A \vert B)= H(A,B)-H(B)$.

\subsubsection{Proving the monogamy inequality for the entropic CHSH}
In order to prove that the monogamy inequality (3) of the main text holds for nonsignaling correlations, we must show that if follows from the elemental inequalities for each subset of mutually compatible observables $A_x,B_y,C_z$. It is sufficient to add the the following elemental inequalities
\begin{eqnarray}
H_{A_0B_0}+H_{A_0C_1} & &\geq H_{A_0B_0C_1}+H_{A_0}, \\
H_{A_0B_1}+H_{A_0C_0} & &\geq H_{A_0B_1C_0}+H_{A_0}, \\
H_{A_1B_0}+H_{B_0C_1} & &\geq H_{A_1B_0C_1}+H_{B_0}, \\
H_{A_1C_0}+H_{B_1C_0} & &\geq H_{A_1B_1C_0}+H_{C_0}, \\
H_{A_0B_0C_1} & &\geq H_{B_0C_1}, \\
H_{A_0B_1C_0} & &\geq H_{B_1C_0}, \\
H_{A_1B_0C_1} & &\geq H_{A_1C_1}, \\
H_{A_1B_1C_0} & &\geq H_{A_1B_1},
\end{eqnarray}
leading to
\begin{eqnarray}
& &H_{A_0B_0}+H_{A_0B_1}+H_{A_1B_0}- H_{A_1B_1} -H_{A_0} -H_{B_0} \\ \nonumber
& & +H_{A_0C_0}+H_{A_0C_1}+H_{A_1C_0}-H_{A_1C_1}-H_{A_0} -H_{C_0} \geq 0,
\end{eqnarray}
that can be rewritten exactly as
\begin{equation}
\label{mono_app}
S^{AB}_{\mathrm{E}}+S^{AC}_{\mathrm{E}}\leq 0,
\end{equation}
with
\begin{equation}
\label{ECHSH_AB_app}
S^{AB}_{\mathrm{E}}=I_{A_0:B_0} +I_{A_0:B_1} +I_{A_1:B_0}-I_{A_1:B_1} -H_{A_0}-H_{B_0},
\end{equation}
and
\begin{equation}
\label{ECHSH_AC_app}
S^{AC}_{\mathrm{E}}=I_{A_0:C_0} +I_{A_0:C_1} +I_{A_1:C_0}-I_{A_1:C_1} -H_{A_0}-H_{C_0}.
\end{equation}

A similar construction can be used to show that the monogamy relation \eqref{mono_app} holds for any symmetry of the entropic CHSH inequalities \eqref{ECHSH_AB_app} and \eqref{ECHSH_AC_app}, thus proving that whenever entropic the marginal correlations $H(A_x,B_y)$ display nonlocality necessarily $H(A_x,C_z)$ must be local.

\subsubsection{Genuine tripartite nonlocality}
Similarly to what happens to entanglement \cite{horodecki2009quantum}, when moving from the bipartite case, different classes of nonlocality can be categorized. In the tripartite scenario one can introduce the notion of genuine tripartite nonlocality, that is, a stronger form of nonlocality that cannot be reproduced by LHV models even if two of the parties are allowed to share some nonlocal resources. As discussed in the main text, hybrid local-nonsignaling (L$|$NS) models are those that can be decomposed as
\begin{eqnarray}
\label{LHV_tri_gen}
p(a,b,c\vert x,y,z) = & & \sum_{\lambda} p(a \vert x, \lambda ) p(b,c \vert y,z, \lambda) p( \lambda ) + \\ \nonumber
& & \sum_{\mu} p(a,b \vert x,y, \mu ) p(c \vert z, \mu) p( \mu ) + \\ \nonumber
& &\sum_{\nu} p(a,c \vert x,z, \nu ) p(b \vert y, \nu) p( \nu ).
\end{eqnarray}
where $\sum_\lambda p( \lambda )+ \sum_\mu p( \mu )+ \sum_\nu p( \nu )=1$ and where  $p(b,c \vert y,z, \lambda)$ (similarly to the other permutations) represent some nonlocal resource. The different nonlocal resources we allow the parties to share will lead to distinct notions of genuine multipartite nonlocality. See \cite{Gallego2012,Pironio2013} for a discussion of the different (non-signalling or signalling) nonlocal resources that can be used in the definition of genuine multipartite nonlocality. Here we will focus our attention to nonsignaling nonlocal resources, as for example, nonlocal quantum correlations or PR-boxes \cite{Popescu1994}. In this case, \eqref{LHV_tri_gen} is equivalent to the existence of probability distributions $p(a_0,a_1,b_j,c_k)$, $p(a_i,b_0,b_1,c_k)$ and $p(a_i,b_j,c_0,c_1)$  such that the marginals $p(a_i,b_j,c_k)$ coincide $\forall i,j,k=0,1$.

To simplify the discussion let us consider that each of the parties can perform two possible measurements. The first term in the decomposition \eqref{LHV_tri_gen}, that is, a model with decomposition given by
\begin{equation}
\label{LHV_tri1}
p(a,b,c\vert x,y,z) = \sum_{\lambda} p(a \vert x, \lambda ) p^{\mathrm{NS}}(b,c \vert y,z, \lambda) p( \lambda ),
\end{equation}
is equivalent, in the entropic description, to the existence of $H(A_0,A_1,B_j,C_k)$ and all its marginals. For each value of $j,k=0,1$ we have therefore a collection of four variables respecting the Shannon type inequalities \eqref{shannonineqs_basic}.

Similarly, the two other terms in \eqref{LHV_tri_gen}, namely
\begin{eqnarray}
\label{LHV_tri2}
p(a,b,c\vert x,y,z) = & & \sum_{\mu} p^{\mathrm{NS}}(a,b \vert x,y, \mu ) p(c \vert z, \mu) p( \mu ), \\\label{LHV_tri3}
p(a,b,c\vert x,y,z) =& & \sum_{\nu} p^{\mathrm{NS}}(a,c \vert x,z, \nu ) p(b \vert y, \nu) p( \nu ),
\end{eqnarray}
imply the existence of $H(A_i,B_0,B_1,C_k)$ and $H(A_i,B_j,C_0,C_1)$ respectively. Thus, following the general prescription, the entropic description of hybrid local-nonsignaling (L$|$NS) correlations corresponds to the intersection of the Shannon cones defined for each of subsets of variables $\left\{A_0,A_1,B_j,C_k\right\}$, $\left\{A_i,B_0,B_1,C_k\right\}$ and $\left\{A_i,B_j,C_0,C_1\right\}$ with $i,j,k=0,1$.

The marginal entropic cone $\Gamma_{\mathrm{L}|\mathrm{NS}}
$ characterizing \eqref{LHV_tri_gen} will be the convex hull of the of the cones $\Gamma^a_{\mathrm{L}|\mathrm{NS}}
$, $\Gamma^b_{\mathrm{L}|\mathrm{NS}}
$ and $\Gamma^c_{\mathrm{L}|\mathrm{NS}}
$, characterizing, respectively, \eqref{LHV_tri1}, \eqref{LHV_tri2} and \eqref{LHV_tri3}. We have tried computationally to compute the extremal rays of $\Gamma_{\mathrm{L}|\mathrm{NS}}
$ however the computation seems to be out of reach. In spite of that, checking whether a given nonlocal extremal ray of $\Gamma^{3}_{\mathrm{NS}}$ defines or not a genuine tripartite nonlocal correlation is computationally much simpler.

Notice that $\Gamma_{\mathrm{L}|\mathrm{NS}}
 \subset \Gamma^{3}_{\mathrm{NS}}$ and since $\Gamma_{\mathrm{L}|\mathrm{NS}}
$ is obtained as the convex hull of $\Gamma^i_{\mathrm{L}|\mathrm{NS}}
$ ($i=a,b,c$), checking whether a given extremal ray of $\Gamma^{3}_{\mathrm{NS}}$ lies inside $\Gamma_{\mathrm{L}|\mathrm{NS}}
$ is equivalent to check whether it lies inside each of the cones $\Gamma^i_{\mathrm{L}|\mathrm{NS}}
$ ($i=a,b,c$). This can be seen as follows.let $r\in \Gamma^{3}_{\mathrm{NS}}$, then $\boldsymbol{r}\in \Gamma_{\mathrm{L}|\mathrm{NS}}
$ if $\boldsymbol{r} =\boldsymbol{h_1} + \boldsymbol{h_2} +\boldsymbol{h_3}$, where $\boldsymbol{h_i} \in \Gamma^i_{\mathrm{L}|\mathrm{NS}}$
Notice that, by cone properties, the sum can be interpreted as a the convex sum $\frac{1}{3}(3\boldsymbol{h}_1 +3\boldsymbol{h}_2 +3\boldsymbol{h}_3)$. This can be written as a linear program (actually, as feasibility problem, i.e., $\boldsymbol{\mathcal{I}}$ is irrelevant)

\begin{eqnarray}
\underset{\boldsymbol{r} \in \RR^{2^n}}{\minimize} & & \langle \boldsymbol{\mathcal{I}}, \boldsymbol{h} \rangle  	\label{LP3} \\
\st & &  M_1\boldsymbol{h_1} \geq \boldsymbol{0}, \\ \nonumber
& &  M_2\boldsymbol{h_2} \geq \boldsymbol{0}, \\ \nonumber
& &  M_3\boldsymbol{h_3} \geq \boldsymbol{0}, \\ \nonumber
& & \boldsymbol{h_1} + \boldsymbol{h_2} +\boldsymbol{h_3} = \boldsymbol{r}.
\end{eqnarray}

In addition, if $\boldsymbol{r}$ is an extremal ray of $\Gamma^{3}_{\mathrm{NS}}$, then
 $\boldsymbol{r} =\boldsymbol{h_1} + \boldsymbol{h_2} +\boldsymbol{h_3}$ implies that at
 least two of the $\boldsymbol{h_i}$ are zero (extremal rays cannot be written as convex
 combinations). Hence, for extremal rays it is sufficient to check that they belong to at
 least one of the $\Gamma^i_{\mathrm{L}|\mathrm{NS}}$.

Thus, running three sets of linear programs, one for each cone $\Gamma^i_{\mathrm{L}|\mathrm{NS}}
$ ($i=a,b,c$), we can decide whether a given extremal entropic NS ray displays or not genuine tripartite nonlocality.

A similar construction holds to check whether a given entropic inequality is a valid witness of GTNL. An entropic inequality $\boldsymbol{\mathcal{I}}$ for GTNL should be satisfied by any correlation that can be written as the convex sum over points within $\Gamma^i_{\mathrm{L}|\mathrm{NS}}
$ ($i=1,2,3$), that is,
\begin{equation}
\label{ineqCs}
\langle \boldsymbol{\mathcal{I}}, \boldsymbol{h}_1 +\boldsymbol{h}_2 +\boldsymbol{h}_3 \rangle \geq 0,
\end{equation}
where $\boldsymbol{h}_i \in \Gamma^i_{\mathrm{L}|\mathrm{NS}}
$.  Any inequality satisfying $\langle \boldsymbol{\mathcal{I}}, \boldsymbol{h}_1 \rangle \geq 0$, $\langle \boldsymbol{\mathcal{I}}, \boldsymbol{h}_2 \rangle \geq 0$ and $\langle \boldsymbol{\mathcal{I}}, \boldsymbol{h}_3 \rangle \geq 0$ will also satisfy \eqref{ineqCs}. This means that in practice to check if a given inequality is a valid GTNL inequality, we need to solve three linear programs like \eqref{LP}, one for each set of inequalities characterizing the cones $\Gamma^i_{\mathrm{L}|\mathrm{NS}}
$ ($i=1,2,3$) .
This can be seen as follows. Consider the LP
\begin{eqnarray}
\underset{\boldsymbol{r} \in \RR^{2^n}}{\minimize} & & \langle \boldsymbol{\mathcal{I}}, \boldsymbol{h_1} + \boldsymbol{h_2} +\boldsymbol{h_3} \rangle  	\label{LP4} \\
\st & &  M_1\boldsymbol{h_1} \geq \boldsymbol{0}, \\ \nonumber
& &  M_2\boldsymbol{h_2} \geq \boldsymbol{0}, \\ \nonumber
& &  M_3\boldsymbol{h_3} \geq \boldsymbol{0}.
\end{eqnarray}
The inequality is valid if and only if $\min \langle \boldsymbol{\mathcal{I}}, \boldsymbol{h_1} + \boldsymbol{h_2} +\boldsymbol{h_3} \rangle \geq 0$. However, since $\boldsymbol{h_i}=\boldsymbol{0}$ is a valid solution for all three cones, the above is equivalent to require separately that $\min \langle \boldsymbol{\mathcal{I}}, \boldsymbol{h_i} \rangle \geq 0$.

We now turn our attention to the entropic inequality (4) of the main text witnessing GTNL and given by
\begin{eqnarray}
\nonumber
\label{EGTNL_app}
& & S_{\mathrm{L}|\mathrm{NS}}
=H_{A_1B_1C_0}+H_{A_1B_0C_0}+H_{A_1B_0C_1}+H_{A_0B_1C_0}+H_{A_0B_1C_1} \\
& &- H_{A_1B_1C_1} -H_{A_1B_0}- H_{A_1C_0}- H_{A_0C_1} - H_{B_1C_0} \geq 0.
\end{eqnarray}

To prove that this inequality is valid for the cone $\Gamma^a_{\mathrm{L}|\mathrm{NS}}
$, it is sufficient to consider the sum of the following basic inequalities (valid for the subsets of variables $\left\{A_0,A_1,B_j,C_k\right\}$):
\begin{eqnarray}
\label{ineq_sum}
H_{A_0A_1B_1C_0}  & & \geq H_{A_0A_1C_0}, \\ \nonumber
H_{A_0A_1B_1C_1}  & & \geq H_{A_1B_1C_1}, \\ \nonumber
H_{A_0A_1B_0C_0}  & & \geq H_{A_0A_1B_0}, \\ \nonumber
H_{A_0A_1B_0C_1}  & & \geq H_{A_0A_1C_1}, \\ \nonumber
H_{A_1B_1C_0} +H_{A_0B_1C_0} & & \geq H_{A_0A_1B_1C_0} + H_{B_1C_0}, \\ \nonumber
H_{A_0A_1C_1} +H_{A_0B_1C_1}  & &\geq H_{A_0A_1B_1C_1} + H_{A_0C_1}, \\ \nonumber
H_{A_0A_1C_0} +H_{A_1B_0C_0}  & &\geq H_{A_0A_1B_0C_0} + H_{A_1C_0}, \\ \nonumber
H_{A_0A_1B_0} +H_{A_1B_0C_1}  & &\geq H_{A_0A_1B_0C_1} + H_{A_1B_0}.
\end{eqnarray}
To prove the validity of \eqref{EGTNL_app} for the cones $\Gamma^b_{\mathrm{L}|\mathrm{NS}}
$ and $\Gamma^c_{\mathrm{L}|\mathrm{NS}}
$ a similar sum of eight basic inequalities is again enough.

As discussed in the main text, the entropic inequality \eqref{EGTNL_app} can be used to witness the presence of genuine tripartite nonlocality in quantum states. To that aim we consider d-dimensional GHZ states
\begin{equation}
\ket{\mathrm{GHZ}}= \frac{1}{\sqrt{d}} \sum_{j=0}^{d-1} \ket{jjj},
\end{equation}
and projective measurements given by \cite{Collins2002}
\begin{equation}
\label{f_measurements}
\ket{k}_{p,m}= \frac{1}{\sqrt{d}}\sum_{j=0}^{d-1} e^{i \frac{2\pi}{d} j (k+\alpha_{p,m})} \ket{j},
\end{equation}
where $p=1,2,3$ denotes the party, $m=0,1$ denotes its measurement choice and $k$ denotes its outcome. For this sort of measurements on GHZ states the probabilities of outcomes can readily be computed to be given by
\begin{equation}
\nonumber
\label{probfourier}
p(a,b,c\vert x,y,z)=  \frac{1}{d^4}\csc^2 \left( \frac{\gamma_{a,b,c,x,y,z}}{d} \right) \sin^2 \left( \gamma_{a,b,c,x,y,z} \right),
\end{equation}
with $\gamma_{a,b,c,x,y,z}=\pi (a+b+c+\alpha_{1,x}+\alpha_{2,y}+\alpha_{3,z})$.

Using this expression we can readily compute the associated Shannon entropies and thus perform numerical optimizations to obtain the optimal violation of inequality \eqref{EGTNL_app}. We have performed such analysis up to $d=40$ and the results are shown in Fig. 2 of the main text.

\subsubsection{Bilocality}
The so-called bilocality scenario \cite{Branciard2010} has been introduced as the classical
analogue of an entanglement swapping experiment \cite{Zukowski1993}, where two independent
pairs of entangled particles are distributed among three parties. Joint measurements on one
particle of each pair (e.g. in a Bell basis) can create entanglement and nonlocal
correlations among the remaining two particles, even though they have never interacted.
Formally, we consider that a central node, Bob, receives one particle from both pairs while
two other parts, Alice and Charlie, receive each one particle from a given entangled pair.
As usual in a Bell scenario, we consider that at each run of the experiment each of the
parties can perform different possible measurements labeled by random variables $X$, $Y$
and $Z$ and obtain, respectively, outcomes labeled by $A$, $B$, $C$. The observed probability distribution $p(a,b,c \vert x,y,z)$
can thus be decomposed as
\begin{eqnarray}
\label{LHV_bilocal}
p(a,b,c \vert x,y,z)= \sum_{\lambda_1,\lambda_2} & &  p(\lambda_1)p(\lambda_2) \\ \nonumber
& &p(a \vert x,\lambda_1 ) p(b \vert y,\lambda_1,\lambda_2 ) p(c \vert z,\lambda_2 ).
\end{eqnarray}
Notice that in \eqref{LHV_bilocal} the finer structure of the underlying causal structure is taken into account, namely in the fact that $p(\lambda_1,\lambda_2)=p(\lambda_1)p(\lambda_2)$ (independence of the sources) and that the outcomes $A$ of Alice ($C$ of Charlie) only depend on $\Lambda_1$ ($\Lambda_2$).

Considering that each of the parties perform two possible measurements, the entropic description of bilocality is equivalent to the existence of a joint entropy $H(A_0,A_1,B_0,B_1,C_0,C_1)$ respecting the Shannon type inequalities together with the bilocality constraint $H(A_0,A_1,C_0,C_1)=H(A_0,A_1)+H(C_0,C_1)$.

We further notice that in the bilocality scenario there are two possible slightly different situations we may want to consider. In the first scenario the random variables $B_0$ and $B_1$ stand for the measurement outcomes of two different measurements performed by Bob. In this case $H(B_0,B_1)$ will correspond to a non-observable quantity. In the second situation, we can understand $B_0$ and $B_1$ as standing for a finer description of a single measurement performed by Bob. For instance, consider that Bob always measure in a Bell basis $\left\{ \ket{\Psi^{\pm}},\ket{\Phi^{\pm}} \right\}$. The variable $B_0$ could stand for the information regarding whether the measurement outcome correspond to $\ket{\Psi}$ or $\ket{\Phi}$ while $B_1$ would stand for the information about phase, for instance,
$\ket{\Psi^{+}}$ or $\ket{\Psi^{-}}$. In this situation, $H(B_0,B_1)$ corresponds to an observable quantity. In the following we will focus our attention to the first case, that is, we have to eliminate from our description terms like $H(B_0,B_1)$.

Similarly to what happens in the usual tripartite case, the complete description of the bilocality scenario in terms of a FM elimination is out of computational reach. However, characterizing the extremal rays of the bilocal entropic NS cone is much simpler. We have obtained $329$ different classes of extremal NS bilocal entropic rays, $15$ of which are bilocal while $314$ correspond to nonbilocal correlations. From this $314$ rays, $40$ are genuine-nonbilocal in the sense of admitting a LHV model but not a bilocal LHV model. Using the tool described in Sec. \ref{subsec:tool} one can derive entropic inequalities detecting the nonbilocality of these rays. Next we employ the framework in Sec. \ref{subsec:tool} to provide an analytical proof for the bilocality inequalities (5) and (6) discussed in the main text.
To prove (5), we have to sum the following basic inequalities
\begin{eqnarray}
H_{A_0}+H_{C} & &\geq H_{A_0C}, \\ \nonumber
H_{A_0A_1BC} & &\geq H_{A_0A_1C}, \\ \nonumber
H_{A_0A_1}+H_{A_0B} & &\geq H_{A_0A_1B} +H_{A_0}, \\ \nonumber
H_{A_0A_1B}+H_{A_1BC} & &\geq H_{A_0A_1BC}+H_{A_1B}.
\end{eqnarray}
and use the bilocality constraint $H_{A_0A_1}+H_{C}=H_{A_0A_1C}$.

In turn, to prove inequality (6) we have to sum the following Shannon type inequalities
\begin{eqnarray}
H_{A_0A_1B_0B_1C_0C_1} & & \geq H_{A_0A_1B_0B_1C_0}, \\ \nonumber
H_{A_0A_1B_0B_1C_0C_1} & & \geq H_{A_0A_1B_0C_0C_1}, \\ \nonumber
H_{B_0C_1}+H_{C_0C_1} & & \geq H_{B_0C_0C_1}+H_{C_1,} \\ \nonumber
H_{A_0A_1}+H_{A_1C_1} & & \geq H_{A_0A_1C_1}+H_{A_1,} \\ \nonumber
H_{A_0B_0C_1}+H_{B_0C_0C_1} & & \geq H_{A_0B_0C_0C_1}+H_{B_0C_1}, \\ \nonumber
H_{A_1B_1C_0}+H_{A_1B_1C_1} & & \geq H_{A_1B_1C_0C_1} +H_{A_1B_1}, \\ \nonumber
H_{A_0A_1C_1}+H_{A_1C_0C_1} & & \geq H_{A_0A_1C_0C_1}+H_{A_1C_1}, \\ \nonumber
H_{A_0B_0B_1C_0}+H_{A_0B_0C_0C_1} & & \geq  H_{A_0B_0B_1C_0C_1}+H_{A_0B_0C_0}, \\ \nonumber
H_{A_1B_0C_0C_1}+H_{A_1B_1C_0C_1} & & \geq H_{A_1B_0B_1C_0C_1} +H_{A_1C_0C_1}, \\ \nonumber
H_{A_0A_1B_0B_1C_0}+H_{A_0B_0B_1C_0C_1} & & \geq H_{A_0A_1B_0B_1C_0C_1}+H_{A_0B_0B_1C_0}, \\ \nonumber
H_{A_0A_1B_0C_0C_1}+H_{A_1B_0B_1C_0C_1} & & \geq H_{A_0A_1B_0B_1C_0C_1}+H_{A_1B_0C_0C_1}. \\ \nonumber
\end{eqnarray}
and use the bilocality constraint $H_{A_0A_1}+H_{C_0C_1}=H_{A_0A_1C_0C_1}$.

\begin{table*}
\label{tab_IC}
  \begin{tabular}{|c|l|l|l|l|l|l|l|}
    \hline
Extremal ray & $H(X_0)$ & $H(X_1)$ & $H(Y_0)$ & $H(Y_1)$ & $H(X_0,Y_0)$ & $H(X_1,Y_1)$ & $H(M)$  \\ \hline
1 & 0 & 0 & 0 & 0 & 0 & 0 & 1 \\ \hline
2 & 0 & 0 & 0 & 1 & 0 & 1 & 0 \\ \hline
3 & 0 & 0 & 1 & 0 & 1 & 0 & 0 \\ \hline
4 & 0 & 1 & 0 & 0 & 0 & 1 & 0 \\ \hline
5 & 1 & 0 & 0 & 0 & 1 & 0 & 0 \\ \hline
6 & 0 & 1 & 0 & 1 & 0 & 1 & 1 \\ \hline
7 & 1 & 0 & 1 & 0 & 1 & 0 & 1 \\ \hline
8 & 1 & 1 & 1 & 1 & 1 & 1 & 1 \\ \hline
\end{tabular}
\caption{All the extremal rays defining the entropic NS cone of the information causality scenario.}
\end{table*}

\subsubsection{Information Causality}
Information causality (IC) \cite{Pawlowski2009} has been introduced as a principle to
explain the degree of nonlocality of quantum mechanics. It basically states that the amount
of information a given part, Bob can have over some bits in possession of another part,
 Alice, is limited by the amount of information communicated from Alice to Bob. Information
causality is respected by quantum correlations, however, as shown in \cite{Pawlowski2009}
any nonlocal correlation stronger then maximum permitted by quantum mechanics, in sense of
surpassing the Tsirelson's bound \cite{Cirel1980quantum} for the CHSH inequality, will
violate it.

IC can be understood as a game between Alice and Bob: Alice receives two independent random bits $X_0$ and $X_1$ and the aim of Bob at each run of the game is to guess the value of one of them, having as resources some pre-shared correlations with Alice and a certain amount of communication sent by her. Which of the two bits he should guess is decided by a random variable $S$. In a classical description we have the variables $X_0,X_1$ representing the input bits of Alice; the variables $Y_0,Y_1$ standing for the guesses of Bob ($Y_s$ is the guess of bit $X_s$ given that $S=s$); $M$ stands for the message sent from Alice to Bob and $\Lambda$ corresponds to the pre-shared correlations between them.

The (classical) entropic description of IC is equivalent to the existence of a joint entropy $H(X_0,X_1,B_0,B_1,M,\Lambda)$ fulfilling the following causal constraints imposed by the rules of the game: $I(X_0:X_1)=0$, $I(X_0,X_1:\Lambda)=0$ and $I(X_0,X_1:B_0,B_1 \vert M, \Lambda)=0$. The first constraint encodes the fact that both bits are uncorrelated and thus have null mutual information (though this is not strictly necessary \cite{Safi2011,Chaves2015a}). The second constraint encodes the fact that the bits received by Alice are independent of the pre-shared correlation with Bob. Finally, the third constraint encodes the fact that conditioned on the message sent by Alice and on the pre-shared correlations, the guesses of Bob should becomes completely uncorrelated with the input bits. A similar description can be given for the case where the pre-shared correlations arise from an entangled state $\rho$ rather than a classical variable $\Lambda$ \cite{Chaves2015a}.

Proceeding with the FM elimination we obtain the description of the marginal entropic cone characterizing the information causality game. To that aim, first one needs to define the marginal scenario of interest. Here, similarly to what has been done in \cite{Pawlowski2009}, we will focus on the marginal scenario $\mathcal{M}_{\textrm{IC}}=\left\{ \left\{ X_0,Y_0 \right\},\left\{ X_1,Y_1 \right\},\left\{ M \right\}  \right\}$. Notice, however, that more general marginal scenarios can be defined \cite{Chaves2015a}. It follows that the only non-trivial entropic inequality describing the IC game (for classical and quantum pre-shared correlations) is given by:
\begin{equation}
\label{IC_ineq_app}
I(X_0:Y_0)+I(X_1:Y_1) \leq H(M).
\end{equation}
This inequality quantifies the qualitative expectation that the amount of information Bob has about Alice's inputs is bounded by the amount of information contained in the message (as quantified by $H(M)$).

We are now in position to define the set of entropic NS correlation for the IC scenario. For the marginal scenario under consideration, the NS set is simply characterized by the Shannon type inequalities $I(X_s:Y_s) \geq 0$, $H(X_s,Y_s) \geq H(X_s)$, $H(X_s,Y_s) \geq H(Y_s)$, $H(M) \geq 0$ together with the constraint $I(X_s:Y_s) \leq H(M)$. The NS cone is described in terms of $8$ extremal rays (shown in the Table \eqref{tab_IC}), $7$ corresponding to correlations that respect the information causality constraint (rays $1$ to $7$) and only $1$ violating it (ray $8$). The extremal ray violating \eqref{IC_ineq_app} is exactly given by the entropic correlations obtained if we replace the pre-shared correlation by a PR-box and apply the protocol discussed in the original IC paper \cite{Pawlowski2009}: Alice inputs $x=x_0\oplus x_1$ in her part of the PR-box, obtaining an outcome $a$ that is then used to encode the message $m=x_0\oplus a$. Bob inputs in his part of the PR-box, $y=0,1$ depending on which bit $X_y$ he is interested and obtains an outcome $b_y$. His guess is given by $g_y=m\oplus b_y=x_0\oplus a\oplus b_y=x_0\oplus y(x_0\oplus x_1)$, that is, he can perfectly guess both outputs of Alice thought only one bit of information ($H(M)=1$) has been sent.

\bibliography{EntNS}

\begin{thebibliography}{75}
\bibitem [{fno({\natexlab{c}})}]{fnote5}%
  \BibitemOpen
  \href@noop {} {} ({\natexlab{c}}),\ \bibinfo {note} {this implies, in
  particular, that for $n \leq 3$ every NS entropic vector has a realization in
  terms of NS probability distribution.}\BibitemShut {Stop}%
\bibitem [{fno({\natexlab{d}})}]{fnote4}%
  \BibitemOpen
  \href@noop {} {} ({\natexlab{d}}),\ \bibinfo {note} {these cones are
  polyhedral because of the polyhedral structure of the entropy cone for two
  and three variables.}\BibitemShut {Stop}%
\bibitem [{fno({\natexlab{e}})}]{fnote2}%
  \BibitemOpen
  \href@noop {} {} ({\natexlab{e}}),\ \bibinfo {note} {this is no longer true
  for affine subspace, which are never used in this context.}\BibitemShut
  {Stop}%
\end{thebibliography}

\end{document}